\documentclass[12pt]{article}
\pdfoutput=1
\usepackage{epsfig}
\usepackage{graphicx}

\usepackage[round,authoryear]{natbib}
\usepackage{upgreek}
\usepackage[pdftex,pdfpagemode={UseOutlines},bookmarks,bookmarksopen,colorlinks,linkcolor={blue},citecolor={blue},urlcolor={red}]{hyperref}

\newcommand{\aap}{A\&A}
\newcommand{\mnras}{MNRAS}
\newcommand{\apj}{ApJ}
\newcommand{\pasp}{PASP}
\newcommand{\aspconf}{ASP Conf.\ Ser.}

\newcommand{\micron}{\,$\upmu$m}

\def\Title#1{\begin{center} {\Large {\bf #1} } \end{center}}

\begin{document}

\Title{Transient Alert Follow-up Planned for CCAT}

\bigskip\bigskip


\begin{raggedright}

{\it Tim Jenness\index{Jenness, T.}\\
Department of Astronomy\\
Cornell University\\
Ithaca 14853, U.S.A.}
\bigskip\bigskip
\end{raggedright}

\section*{Abstract}

CCAT is a sub-millimeter telescope to be built on Cerro Chajnantor in
Chile near the ALMA site. The remote location means that all observing
will be done by remote observers with the future goal of fully autonomous
observing using a dynamic scheduler. The fully autonomous observing
mode provides a natural means for accepting transient alert notifications for
immediate follow up.

\section{Introduction}

\begin{figure}
\includegraphics[width=\textwidth]{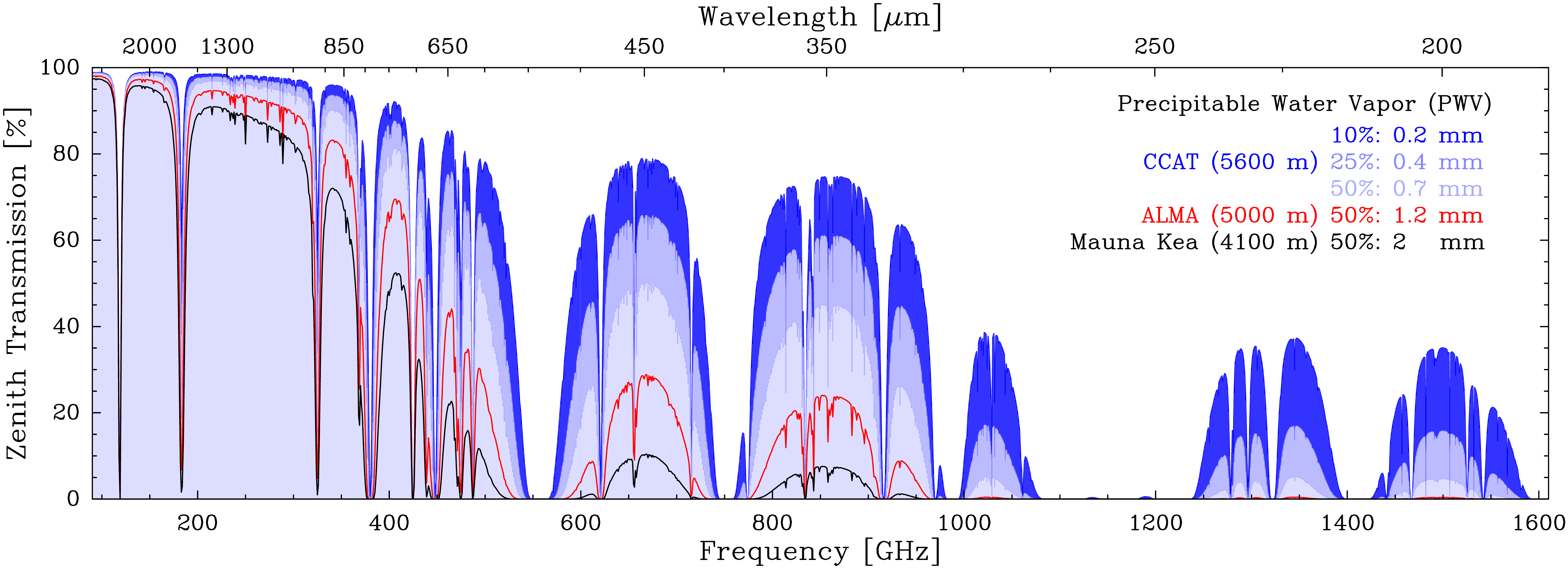}
\caption{Atmospheric transmission for exceptional (10\%), excellent
  (25\%), and median conditions at CCAT and for median conditions at
  ALMA and at Mauna Kea \citep[ATM model;][]{2001ITAP...49.1683P}. Water
  vapor (PWV) distributions determined from 350\micron\ tipper
  measurements \citep{2011RMxAC..41...87R}. (figure credit: S.\
  Radford).}
\label{fig:trans}
\end{figure}

CCAT \citep{2012SPIE.8444E..2MW,2013AAS...22115006G,P10_adassxxiii} is
a 25\,m diameter sub-millimeter telescope to be built on Cerro
Chajnantor in Chile near the ALMA site at an altitude of 5600\,m. The
additional height above the ALMA array results in significant
improvements in transparency across all observing bands (Fig.\
\ref{fig:trans}), and 1.64 times better than the ALMA site at 350\,\micron\
\citep{2011RMxAC..41...87R}. CCAT will initially operate from
350\micron\ to 2\,mm but will be capable in the future of operating at
200\micron\ in the very best weather.

The CCAT project has identified four major first generation
instruments to achieve its science goals. SWCam
\citep{2013AAS...22115007S} will be the first-light camera having of
order 60,000 detectors operating mainly at 350\micron\ with additional
detectors out to 2\,mm. CHAI \citep{GoldsmithCHAI2012} will be a large
format heterodyne array operating in two bands with the backend able
to process spectra with a bandwidth of 4\,GHz and 64,000 channels.
LWCam \citep{2013AAS...22115008G} is a dedicated long-wave camera
operating in 5-6 bands between 750\,\micron\ and 2.1\,mm with a
long-wavelength goal of 3.3\,mm. X-Spec \citep{2013AAS...22115009B} is
a multi-object spectrometer with $\sim$\,100 beams on the sky, each
covering a frequency range of 190-520\,GHz in two bands simultaneously
with a resolving power of 400 -- 700.

\section{The Case for Transient Follow up}

In the sub-millimeter, variable sources are sometimes monitored regularly as
part of general observatory operations of flux calibrators
\citep[e.g.][]{2002MNRAS.336...14J} and of pointing sources such as
blazars \citep{2010MNRAS.401.1240J}. Bright, time-varying sources do
not generally require the ability to respond rapidly to time-sensitive
alerts and can be observed as part of a monitoring program or as a
general target of opportunity.  Detecting the afterglows of gamma ray
bursts (GRBs) in the sub-millimeter has proven to be difficult with the current
generation of instrumentation \citep[see e.g.][]{2012A&A...538A..44D}
and the sooner that a telescope can get on target the more chance
there is to see the peak of the light curve in the sub-millimeter.
GRB\,120422A \citep{2014arXiv1401.3774S} failed to
detect any emission in the sub-millimeter despite being on source within 45
minutes and observing for nearly 2 hours with SCUBA-2
\citep{2013MNRAS.430.2513H}. GRB\,130427A, the brightest GRB in nearly
30 years \citep{2014ApJ...781...37P}, was not observed in the sub-millimeter but
radio data and modeling suggests that the 850\,$\mu$m flux would have
been approximately 1~mJy after 2 days but more than 10~mJy if it had
been observed within 4 hours of detection. First generation CCAT
instruments such as SWCam will be able to observe an area of 0.15\,sq~deg
to a depth of 1 mJy in only an hour in good weather. This is
significantly better performance than current sub-millimeter instrumentation
and indicates that the chances of detecting GRBs will increase
considerably.

In addition to GRBs, LSST \citep{2008arXiv0805.2366I} will be coming
online at around the same time as CCAT and will begin publishing
millions of alerts per night. Some of these will be of interest to
sub-millimeter astronomers and require reasonably fast follow up observations.

Once instrumentation has sufficient sensitivity to be useful, the main
issue associated with time-sensitive alerts is how to respond to them
in a timely manner. This is especially important for a common-user
telescope designed for survey and P.I.\ observations.

\section{Reacting to Alerts}

The CCAT observation scheduler will initially be a dynamic JIT (`just
in time') scheduler determining the best observation block to observe
at the current time. The system will be similar to that used by ALMA
\citep{2007ASPC..376..673L} and the James Clerk Maxwell Telescope
\citep{2002ASPC..281..488E}. A human operator, based either in San
Pedro or at a remote observing location, will use the scheduler to
guide the observing program and make the final choice of targets and
associated calibrations.

\begin{figure}[t]
\includegraphics[width=\textwidth]{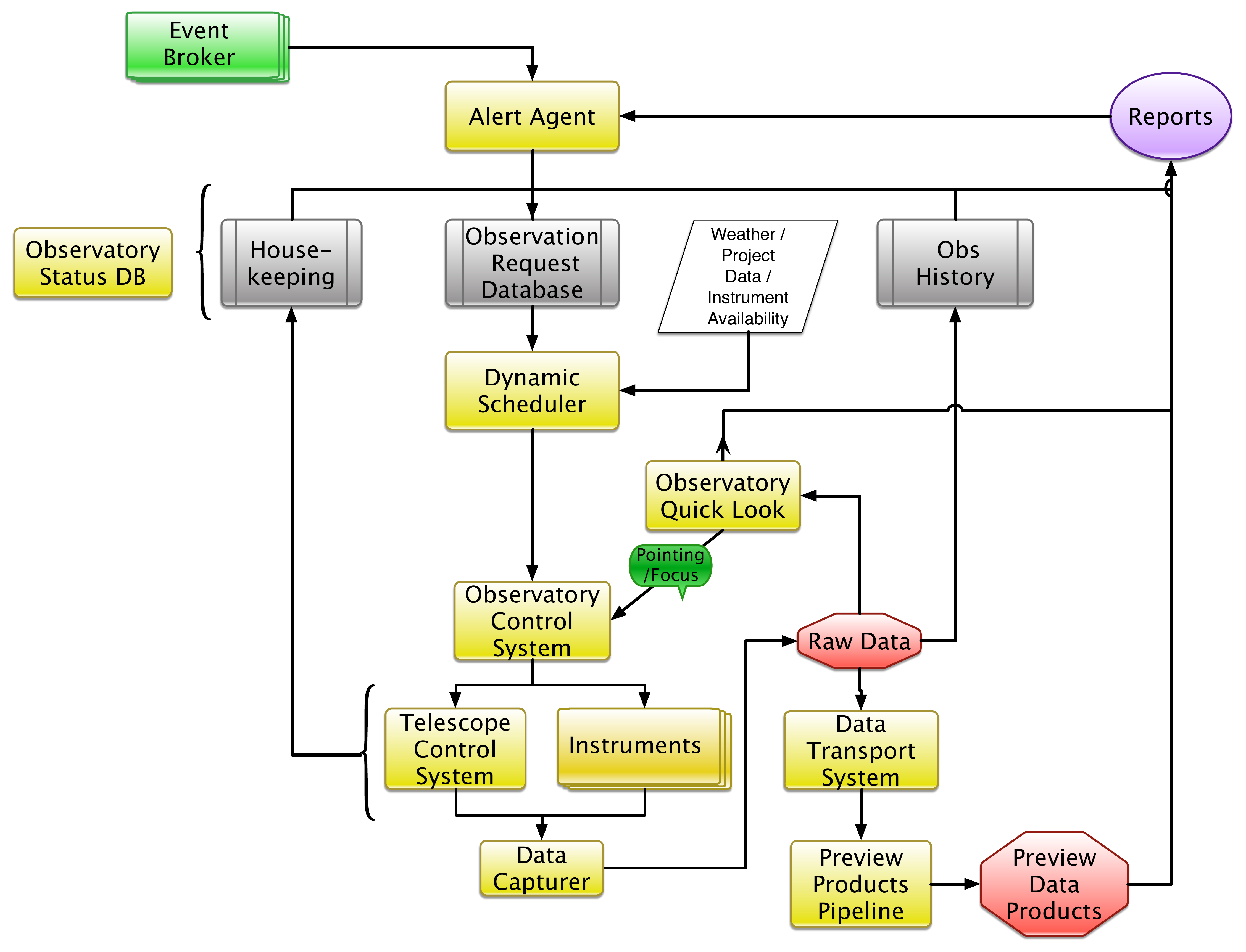}
\caption{Flow chart of the CCAT software system from the perspective
  of processing transient alerts.}
\label{fig:system}
\end{figure}

The infrastructure being designed as part of the observation
management system provides easy programmatic interface to the
observation request database. The system architecture is shown in Fig.\
\ref{fig:system}. The concept is that an alert broker, for example
something like the ANTARES broker \citep{2014AAS...22334302M,HWTUIII_Matheson}, will
send a VOEvent message \citep[e.g.][]{2006ASPC..351..637W}
to an alert agent. The alert agent will be run
by an interested astronomer, possibly at their home institution. If
the alert is of interest an observation will be submitted to the
observation request database at the telescope.  Once this minimum
schedulable block has been submitted to the database the standard
system will be used and the data will be processed in the normal
way. Quality Assurance information and, possibly, flux measurements,
will be fed back to the alert agent to allow the astronomer or agent
to schedule follow up observations automatically.

This design is similar to that implemented at the UKIRT telescope
\citep{2006AN....327..788E,2011tfa..confE..42J} which responded to a
GRB alert within a few minutes \citep{2009GCN..9202....1T} using the
eSTAR system \citep{2004SPIE.5496..313A}.

\section{Post Commissioning}

The goal, following telescope commissioning of the base system, is to
upgrade the scheduler to fully autonomous operation \citep[see
e.g.][for background]{LampoudiS13} where the observing queue will be monitored
continuously and observations submitted as needed, calibrations will
be scheduled when appropriate and observation blocks will be accepted
or rejected automatically based on quality assurance data from the
instrument pipelines. In the sub-millimeter it is sometimes the case
that a flux calibrator will not be available until later in the night
and so care must be taken to keep track of calibration data that are
required for observations that have already been taken. This
scheduling ability would allow would allow time-sensitive followups to
be inserted directly into the queue and observed without human
intervention, similar to a fully robotic telescope such as LCOGT
\citep{2013PASP..125.1031B,2010SPIE.7737E..17H}.

\bigskip

\paragraph{Acknowledgments}

The CCAT Submillimeter Observatory (CCAT) is owned and operated by a
consortium of universities and non-profit organizations located in the
United States, Canada and Germany. Specifically the CCAT Consortium is
comprised of: Cornell University, California Institute of Technology
(Caltech), University of Colorado at Boulder, University of Cologne,
University of Bonn, Dalhousie University, McGill University, McMaster
University, University of British Columbia, University of Calgary,
University of Toronto, University of Waterloo, University of Western
Ontario and Associated Universities, Incorporated.  The CCAT
Engineering Design Phase was partially supported by funding from the
National Science Foundation via AST-1118243.

\end{document}